\documentclass[
 reprint,
 amsmath,amssymb,
 aps,
longbibliography
]{revtex4-1}

\usepackage{graphicx}
\usepackage{dcolumn}

\usepackage{natbib}
\setcitestyle{authoryear,round}

\usepackage{hyperref}

\usepackage{amsmath} 
\usepackage{bbm} 

\renewcommand{\say}[1]{``#1''}

\begin{document}

\title{Solving the wave equation with physics-informed deep learning}

\author{B. Moseley}
\email{bmoseley@robots.ox.ac.uk}
\author{A. Markham}

\affiliation{Department of Computer Science
\\ University of Oxford}

\author{T. Nissen-Meyer}

\affiliation{Department of Earth Sciences
\\ University of Oxford}




\date{\today}

\begin{abstract}
We investigate the use of Physics-Informed Neural Networks (PINNs) for solving the wave equation. Whilst PINNs have been successfully applied across many physical systems, the wave equation presents unique challenges due to the multi-scale, propagating and oscillatory nature of its solutions, and it is unclear how well they perform in this setting. We use a deep neural network to learn solutions of the wave equation, using the wave equation and a boundary condition as direct constraints in the loss function when training the network. We test the approach by solving the 2D acoustic wave equation for spatially-varying velocity models of increasing complexity, including homogeneous, layered and Earth-realistic models, and find the network is able to accurately simulate the  wavefield across these cases. By using the physics constraint in the loss function the network is able to solve for the wavefield far outside of its boundary training data, offering a way to reduce the generalisation issues of existing deep learning approaches. We extend the approach for the Earth-realistic case by conditioning the network on the source location and find that it is able to generalise over this initial condition, removing the need to retrain the network for each solution. In contrast to traditional numerical simulation this approach is very efficient when computing arbitrary space-time points in the wavefield, as once trained the network carries out inference in a single step without needing to compute the entire wavefield. We discuss the potential applications, limitations and further research directions of this work.
\end{abstract}

\maketitle


\section{\label{sec:introduction}Introduction}

Waves are the most effective means of information transfer over many distance scales, and they play a central role in natural as well as technological phenomena. Solving the wave equation is therefore a hugely important task for a wide range of physics problems. It can take many forms and arises in fields such as acoustics, electromagnetics, cosmology, fluid dynamics and not least in the field of geophysics, where it is needed to simulate earthquakes \citep{Boore2003, Cui2010}, estimate subsurface structure \citep{Tarantola1987, Schuster2017, Fichtner2010, Virieux2009}, carry out non-destructive testing \citep{Liu2005}, model tsunamis \citep{Maeda2013} and characterise the interior of the Earth and other planets \citep{Hosseini2019, Stahler2018}, amongst many other applications.

The most popular methods for solving the wave equation are Finite Difference (FD) and spectral element methods, which are general methods for solving differential equations \citep{Igel2017, Moczo2007, Komatitsch1999, Leng2019}. In these approaches the wave equation is discretised and iterative time-stepping schemes are used to solve it. These methods have benefited from decades of development, and highly sophisticated algorithms exist which are able to model a large range of physics; for example in geophysics, the effects of solid-fluid interfaces, intrinsic attenuation and anisotropy can be readily incorporated into the solution \citep{Fernando2020, Colombi2012, VanDriel2014, Tesoniero2020}.

However, a key difficulty when using these methods is their computational cost. For large 3D solutions, billions of mesh points may be required needing large supercomputers to run \citep{Bohlen2002, Leng2019}. Furthermore, they suffer from discretisation error and this typically requires the grid size to be made much smaller than the rate of change of the system for it to be acceptable \citep{Courant1967}, especially if the waves are propagating over thousands of wavelengths of distance \citep{Nissen-Meyer2014}. Unless local time stepping is used, temporal sampling of the dynamic evolution often needs to be extremely small to accommodate the constraints imposed by the meshing.

In recent years modern deep learning techniques have shown excellent promise in their ability to simulate physical phenomena. For example, \cite{Paganini2018} showed deep neural networks could simulate particle showers in particle colliders, \cite{Rasp2018} showed they could provide fast emulators for subgrid processes in climate simulations and \cite{Kasim2020} proposed a general neural architecture search which was able to accelerate simulations across a wide range of scientific tasks. Deep learning is also having a large impact in the field of fluid mechanics \citep{Brunton2020}. For solving the wave equation, \cite{Moseley2020} showed that deep neural networks can simulate waves in complex faulted media and \cite{Siahkoohi2019} showed that FD simulation can be accelerated by using  convolutional neural networks and larger time steps. However, a key issue with many of the existing approaches above is their inability to generalise outside of their training distribution \citep{Moseley2020, Zhang2018}. These networks typically rely entirely on their training data and therefore perform poorly outside of it. Generalising outside of the training data is a challenge of deep neural networks in general and is important to overcome if these \say{naive} methods are to move on to more realistic applications \citep{Moseley2020}.

A recent development is the rise of physics-informed machine learning \citep{Karpatne2017a, Arridge2019, Raissi2019}. Instead of replacing known physics with purely data-driven deep neural networks in a wholesale fashion, the idea is to blend physical constraints into the workflow in an attempt to combine the best of both worlds. In particular, a recent approach suggested by \cite{Raissi2019} proposes physics-informed neural networks (PINNs). They design a general scheme for solving the differential equations governing a physical system where a neural network represents a solution of the system. Importantly, the underlying equations are used to directly constrain the network whilst training. \citeauthor{Raissi2019} showed that these networks could learn solutions which were far outside of the boundary data used to train them, which offers a potential way to address the generalisation issues of deep neural networks.

The introduction of PINNs has prompted a large amount of follow up work \citep{Yang2020, Jagtap2020, Yang2019}. For example, \cite{SahliCostabal2020} used them to diagnose atrial fibrillation, \cite{Chen2020} used them to solve inverse scattering problems in nano-optics, and there has been much investigation in their use for fluid mechanics \citep{Zhu2019b, Sun2019a, Raissi2020, Erichson2019}. In geophysics \cite{Xu2020} and \cite{CostaNogueiraJunior2019} used PINNs to constrain seismic inversion, \cite{Shukla2020} used them for ultrasound inversion and \cite{Zhang2020} used them for structural response testing. \cite{Smith2020} presented a network to predict the travel times of waves in heterogeneous media by solving the Eikonal equation. However, little work has been carried out on their use with the wave equation.

Compared to the equations already studied, the wave equation presents unique challenges because of the multi-scale, propagating and oscillatory nature of its solutions, and it is unclear how well PINNs perform in this setting. In this work we investigate the effectiveness of physics-informed neural networks for solving the wave equation. Whilst we place emphasis towards its applications in geophysics, we believe our results are informative for solving the wave equation in general. Our contribution is as follows;
\begin{itemize}
\itemsep -0.1em
\item[--] We design a physics-informed neural network for solving the wave equation and show it is able to accurately solve the pressure response in complex 2D acoustic media. The network is only given the first few timesteps of the solution as training data and is able to accurately predict the solution at much later times. It is able to model a wide range of physical phenomena induced by the wave equation, including the transmission and reflection of waves, the compression and expansion of waves through different velocity regions and the attenuation of waves due to spherical divergence.
\item[--] We extend the original PINN approach proposed by \cite{Raissi2019} by conditioning the network on the initial source location, showing that it is able to generalise over this initial condition without needing to retrain the network. This allows this network to be much more efficient at computing the solution for varying source locations compared to traditional FD modelling.
\item[--] We propose a curriculum-learning based strategy for training the PINN, rather than the standard approach used by \cite{Raissi2019}, which improves convergence.
\item[--] We find and discuss specific challenges when using PINNs to solve the wave equation, such as dealing with discontinuities at the interfaces of media, and highlight ways to address them.
\end{itemize}

Compared to \cite{Smith2020} we solve the wave equation, rather than the Eikonal equation, and compared to \cite{Xu2020} and \cite{CostaNogueiraJunior2019} our focus is on simulation using the wave equation rather than inversion. \cite{Sitzmann2020} presented a physics-informed network for solving the wave equation, although they only considered a constant velocity model. We extend the original approach proposed by \cite{Raissi2019} by conditioning the network on its initial conditions and by proposing a curriculum learning strategy.

In Section~\ref{sec:methods} we describe PINNs and our workflow using them to solve the wave equation. In Section~\ref{sec:results} we test our approach by simulating 2D waves in acoustic media of varying complexity and in Section~\ref{sec:discussion} we discuss the potential applications, limitations and further research directions of this work.

\section{\label{sec:methods}Methods}

\begin{figure}[t]
\begin{center}
\includegraphics[width=8cm]{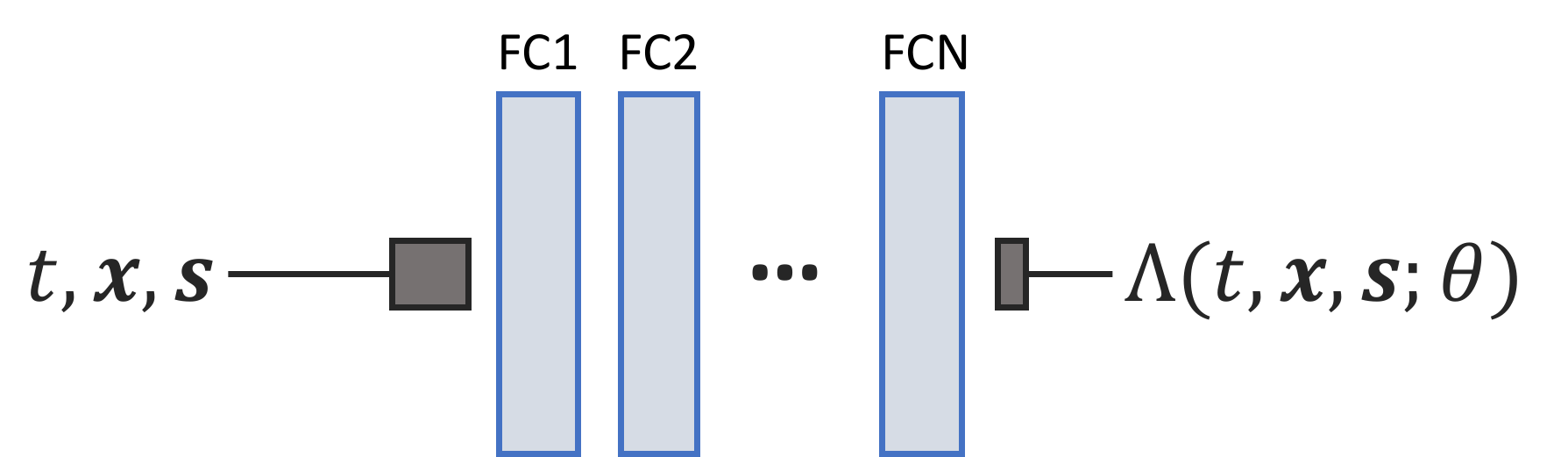}
\caption[]{Physics-informed neural network used to solve the wave equation. The input to the network is a single point in time and space $(t,x)$ and its output is an approximation of the wavefield solution at this location. We extend the original PINN approach proposed by \cite{Raissi2019} by further conditioning the network on the initial source location $s$. We use a fully connected network architecture with 10 layers, 1024 hidden channels, softplus activation functions before all hidden layers and a linear activation function for the final layer.}
\label{fig:model}
\end{center}
\end{figure}

\begin{figure*}[t]
\begin{center}
\includegraphics[width=16cm]{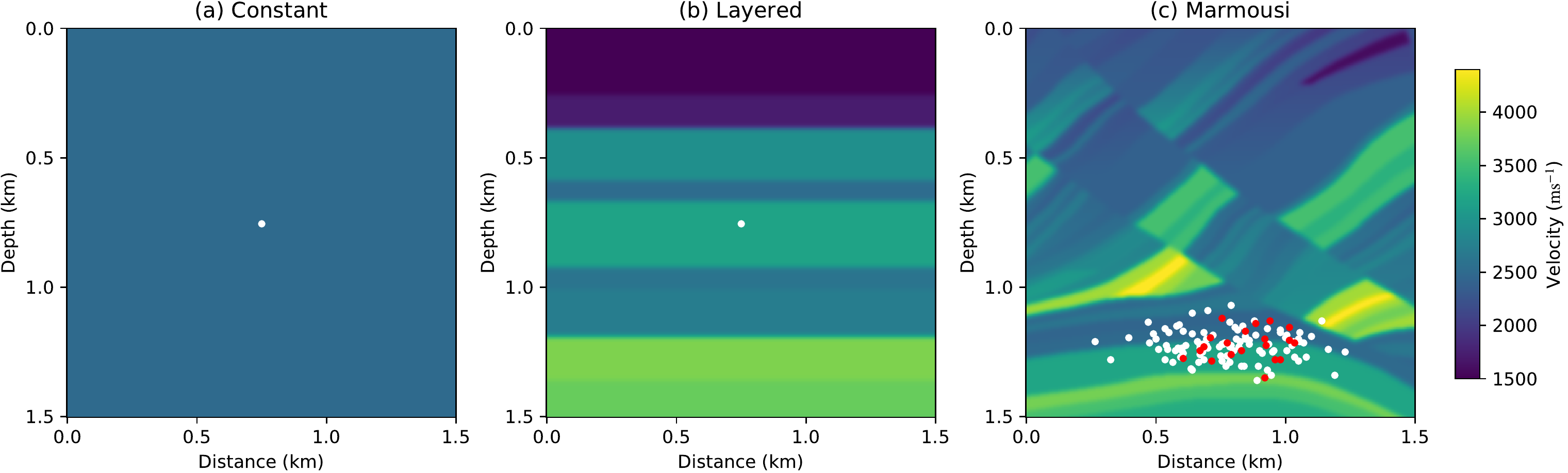}
\caption[]{Velocity models used to test our approach. We use a PINN to solve the wave equation in a variety of different media, starting from a simple homogenous velocity model (left), to a layered velocity model (middle) and finally a region of the Earth-realistic Marmousi model \citep{Martin2006} (right). The velocity model is an implicit parameter in the training scheme and the network must be retrained to solve the wave equation for each case. White points show the locations of the sources used to train the network in each case. For the Marmousi model we train the network to generalise over different source locations and red points show the source locations used to test the network.}
\label{fig:velocity}
\end{center}
\end{figure*}

\subsection{Physics-informed neural networks}

PINNs attempt to solve physical systems which can be written as  
\begin{equation}
\label{eq:rassi}
\mathcal{N}[u(t,x);\lambda]=0~,~x \in \mathbbm{R}^{\mathrm{D}},~t \in \mathbbm{R}~,
\end{equation}
where $\mathcal{N}[u(t,x);\lambda]$ is an underlying differential operator describing the physical system, parameterised by $\lambda$, and $u(t,x)$ represents the solution of the system \citep{Raissi2019}. Many different physical phenomena can be described in this form, including conservation laws, equations of motion, fluid dynamics and many forms of the wave equation. 

The basic idea of PINNs is to use a neural network $\Lambda(t,x;\theta)$, parameterised by $\theta$, to approximate the solution of the physical system $u(t,x)$ and to use Equation~\ref{eq:rassi} as a constraint when training the network. Importantly, the network is a direct functional approximation of $u(t,x)$. It is trained using a loss function which includes both a boundary condition and the underlying equation, given by
\begin{multline}
\label{eq:rassi_loss}
\mathcal{L}={1 \over N_{u}} \sum^{N_{u}}_{i=1}\|u(t_{i},x_{i})-\Lambda(t_{i},x_{i})\|^{2}+ \\
{1 \over N_{\Lambda}} \sum^{N_{\Lambda}}_{j=1}  \|\mathcal{N}[\Lambda(t_{j},x_{j};\theta);\lambda]\|^{2}~,
\end{multline}
where $u(t_{i},x_{i})$ are known initial values of the solution at some boundary points $(t_{i},x_{i})$, and $(t_{j},x_{j})$ are points sampled from the entire input space. By constructing the loss function in this way, the first term (denoted the \say{boundary loss}) attempts to ensure the network learns a unique solution and the second term (denoted the \say{physics loss}) tries to ensure that the network honours the underlying equation. A notable benefit of this approach is that it does not require discretisation, in contrast to traditional numerical methods. From a machine learning point of view, the second term in the loss function can be seen as an unsupervised regulariser and thus the network is likely to have better generalisation performance outside of the boundary data than one trained with just the boundary loss.

It is important to note that the gradients of neural networks with respect to their inputs are typically analytically available and therefore the physics loss in Equation~\ref{eq:rassi_loss} can be derived. Furthermore, a key practical enabler in this approach is the development of auto-differentiation in modern deep learning packages \citep{Baydin2017, tensorflow, pytorch}, which allows such gradients to be easily derived, and to be further differentiated with respect to the network weights $\theta$ in order to train the network.

\subsection{Solving the wave equation}

In this work we investigate the use of PINNs for solving the wave equation. To test our approach, we focus on the 2D acoustic wave equation, given by
\begin{equation}
\label{eq:waveeq_full}
\rho \nabla \cdot \left( {1 \over \rho } \nabla u \right) - {1 \over v^{2} } {\partial^{2} u \over \partial t^{2} } = - \rho {\partial^{2} f \over \partial t^{2} }~,
\end{equation}
where in this setting $u(t,x)$ represents the pressure response in an acoustic medium (known as the \say{wavefield}), $f(t,x)$ is a point source of volume injection, $v = \sqrt{\kappa / \rho}$ is the velocity of the medium, $\rho$ is the density of the medium and $\kappa$ is the adiabatic compression modulus \citep{Long2013}. In general both the density and velocity of the medium can vary spatially. When the density of the medium is constant and the source term is negligible Equation~\ref{eq:waveeq_full} reduces to the canonical form of the wave equation, given by
\begin{equation}
\label{eq:waveeq}
\nabla^{2} u - {1 \over v^{2} } {\partial^{2} u \over \partial t^{2} } = 0 ~.
\end{equation}
Whilst we concentrate on the acoustic wave equation here, we note that our approach is general and is readily extendable to more complex forms.

Despite its linearity, the wave equation is notoriously difficult to solve in complex media, not least due to the broadband, oscillatory and dispersive nature of its solutions. The dynamics of the wavefield can be highly complex, and include reflected, transmitted and grazing waves at interfaces in the media (including prolonged \say{bouncing} around of the wavefield between interfaces), wavefront compression, expansion and healing through different velocity regions, spherical spreading, multiple types of waves interfering simultaneously and a large range of amplitudes and frequencies \citep{Igel2017}. Compared to some of the nonlinear differential equations already investigated by PINNs, such as those in fluid mechanics, it is by no means easier to solve, and it is unclear how well PINNs perform in this setting.

From a machine learning perspective, a number of specific challenges arise when applying PINNs to the wave equation. Firstly, it is uncertain whether neural networks are expressive enough to represent the full range of dynamics stated above. Secondly, the wave equation is a second order partial differential equation and requires strict boundary conditions on both the initial wavefield and its derivatives for its solution to be unique; it is unclear if the boundary loss in Equation~\ref{eq:rassi_loss} will be sufficient in this setting. Thirdly, from a computational point of view the second order gradients of a network are typically much more expensive to compute than its first order derivatives and this potentially restricts the type and size of neural networks which can be used.

In this work we address the first point by testing the ability of PINNs to solve the wave equation in a variety of different conditions, starting from simple solutions in homogeneous media to complex dynamics in heterogeneous Earth-realistic media. We also introduce a curriculum learning-based strategy to help train the network. For the second point, we include multiple time steps in the boundary training data to help constrain both the temporal and spatial gradients of the initial solution and for the third point we restrict ourselves to a relatively small 10 layer fully connected network where its second order gradient computations with respect to its inputs are tractable. We also use a continuous softplus activation function, rather than the more standard piece-wise linear ReLU activation function, to ensure that the network has non-zero second order derivatives.

A limitation of the original PINN approach is that the network must be retrained to solve every new initial condition of the physical system. This is practically undesirable, as it is likely to take as long to train the network as it takes to solve the underlying equations using traditional methods, or longer, at least for the wave equation. In this work, for our last case study (the Earth-realistic media), we condition the network on the initial source location  and train over many different source locations, allowing it to generalise over this initial condition without needing to be retrained. We now discuss our network design and training scheme in detail.

\begin{figure*}[t]
\begin{center}
\includegraphics[width=14cm]{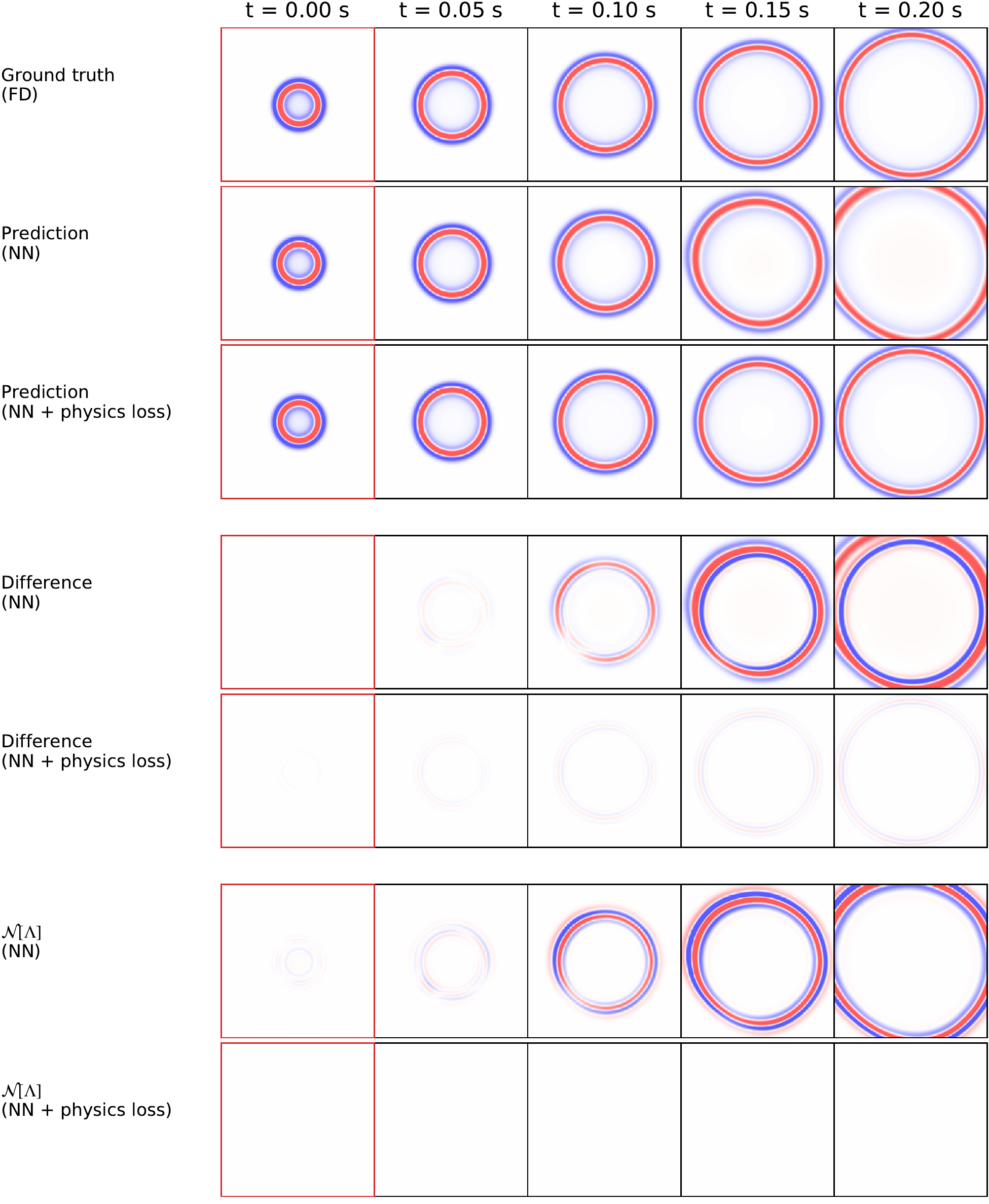}
\caption[]{Comparison of the PINN wavefield prediction to ground truth FD simulation, using a homogeneous velocity model. The PINN (\say{NN + physics loss}) is also compared to the same network trained using only the boundary loss (\say{NN}). Top three rows show the FD, PINN and uninformed network wavefield solutions through time. Middle two rows show the difference of the two network predictions to FD simulation. Bottom two rows show the value of the differential operator in the physics loss in Equation~\ref{eq:loss} (right hand term) for both networks, which is close to zero for an accurate solution to the wave equation. Plots bordered in black indicate wavefield times which are outside of the boundary training data. The boundary training data only covers the time range 0.00-0.02~s and the PINN is able to generalise to much later times (beyond 0.20~s, 10$\times$ the range of the boundary data). Colour bar ranges are the same between each type of plot, for fair comparison. The colour bar is shown in Figure~\ref{fig:gradients}.}
\label{fig:constant}
\end{center}
\end{figure*}

\begin{figure*}[t]
\begin{center}
\includegraphics[width=14cm]{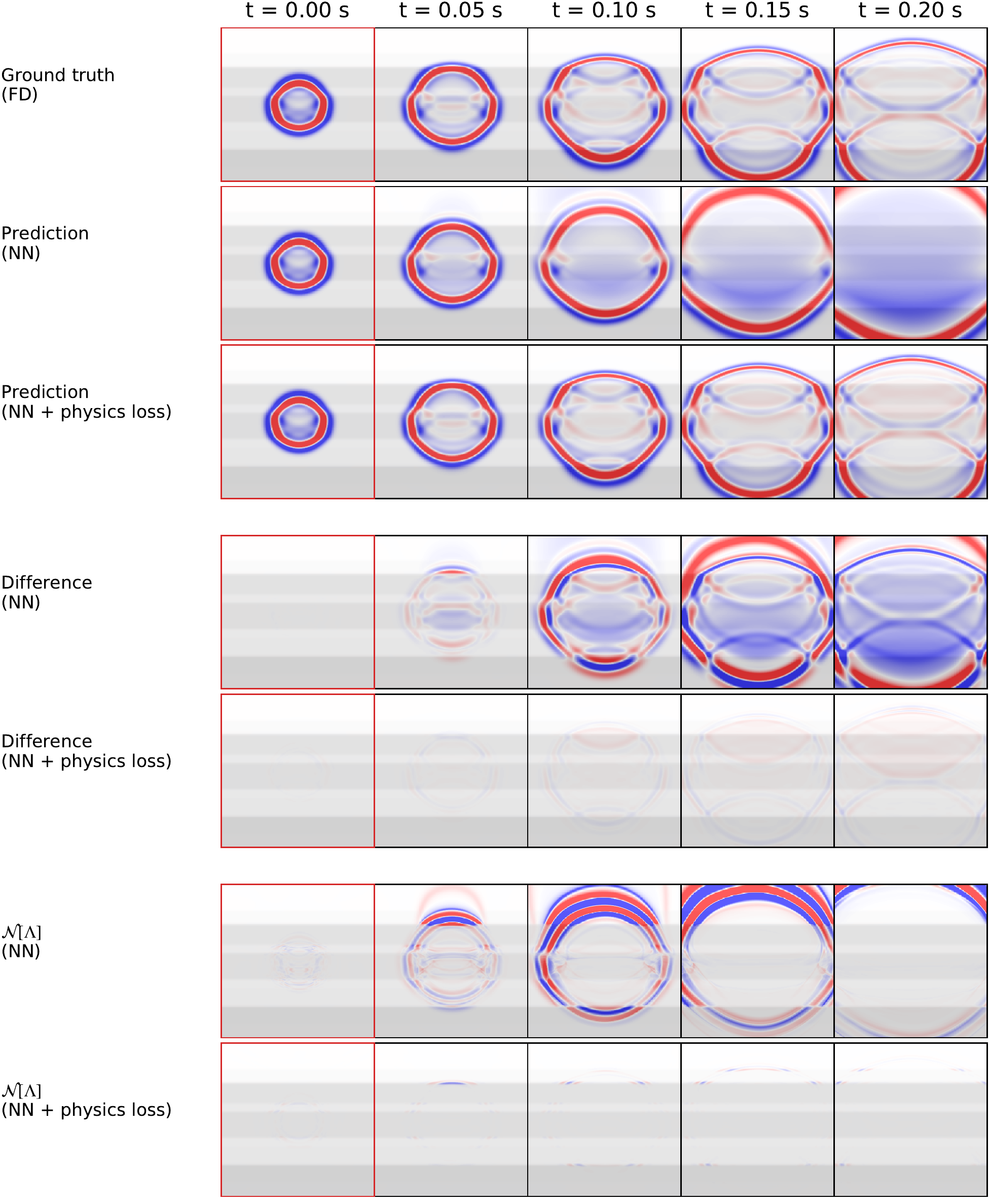}
\caption[]{Comparison of the PINN wavefield prediction to ground truth FD simulation, using a layered velocity model. The same layout as Figure~\ref{fig:constant} is used; top three rows show the FD, PINN and uninformed network wavefield solutions through time; middle two rows show the difference of the two network predictions to FD simulation; bottom two rows show the value of the differential operator in the physics loss in Equation~\ref{eq:loss} (right hand term) for both networks, which is close to zero for an accurate solution to the wave equation; plots bordered in black indicate wavefield times which are outside of the boundary training data. In this case the plots are overlain on the velocity model, shown in grey. Similar to Figure~\ref{fig:constant} the boundary training data only covers the time range 0.00-0.02~s and the PINN is able to generalise to much later times (beyond 0.20~s, 10$\times$ the range of the boundary data).}
\label{fig:layer}
\end{center}
\end{figure*}

\begin{figure*}[t!]
\begin{center}
\includegraphics[width=13cm]{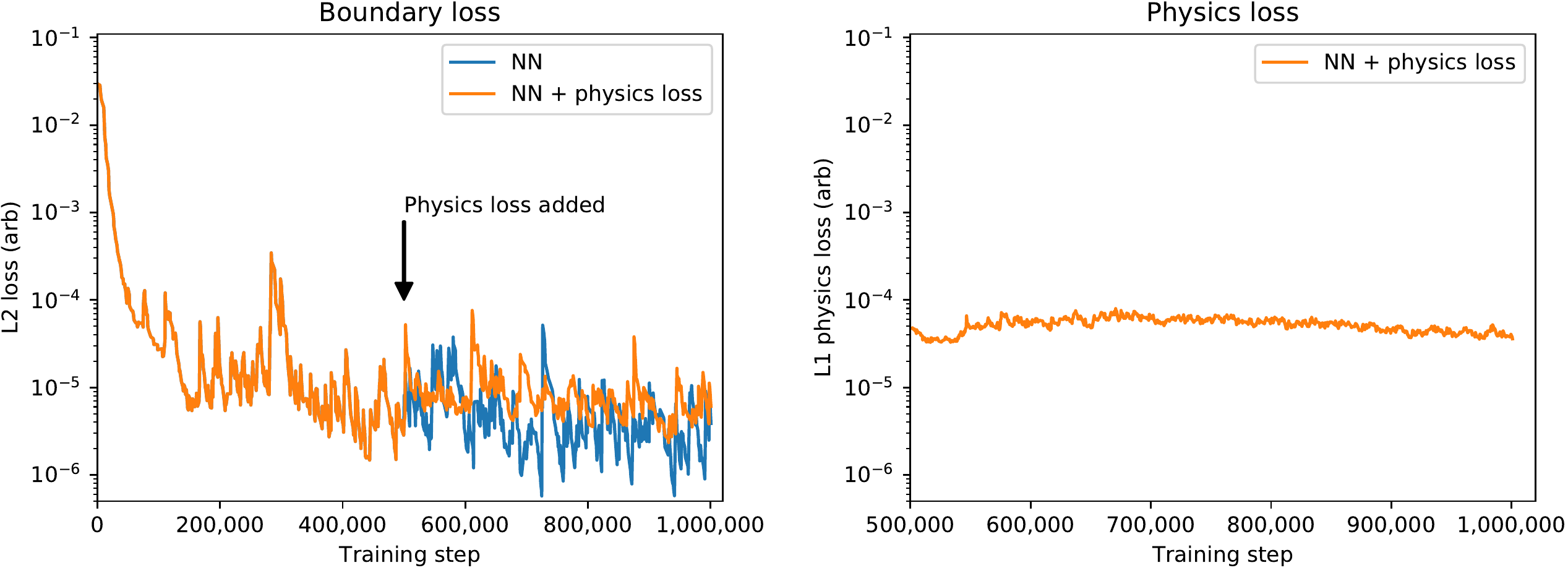}
\caption[]{Training loss for the layered velocity model against training step. Left shows the boundary loss and right shows the physics loss in Equation~\ref{eq:loss}. The physics loss is added halfway through training and includes an expanding time horizon to help the network converge. The PINN (\say{NN + physics loss}) is compared to a network trained only using the boundary loss (\say{NN}).}
\label{fig:loss}
\end{center}
\end{figure*}

\subsection{PINN design and training strategy}

Our PINN workflow used for solving the wave equation is shown in Figure~\ref{fig:model}. The input to the network is an arbitrary point in time and space $(t,x)$ and its output is an approximation of the wavefield solution at this location. A fully connected network architecture is used with 10 layers, 1024 hidden channels, softplus activation functions before all hidden layers and a linear activation function for the final layer. The network is designed to solve the wave equation in a fixed media (implicitly defined in the physics loss) and must be retrained for each new media.

We use data from FD simulation as a boundary condition to train the network. Given a fixed velocity model, density model, source location and source signature as initial conditions, we run FD simulation for a small number of starting time steps to obtain an initial (discretised) wavefield. Values from this initial wavefield $u_{\mathrm{FD}}(t_{i},x_{i}),~x_{i} \in [0, X_{\mathrm{max}}],~t_{i} \in [0,T_{1}]$ are used to compute the boundary loss in Equation~\ref{eq:rassi_loss}. Importantly, only a small number of initial timesteps ($t_{i} < T_{1}$) are used to train the network. For the physics loss in Equation~\ref{eq:rassi_loss}, we randomly sample points $(t_{j},x_{j}),~x_{j} \in [0, X_{\mathrm{max}}],~t_{j} \in [0,T_{\mathrm{max}}],~T_{\mathrm{max}} \gg T_{1}$ over a much larger input space, which includes times much later than the initial wavefield, to attempt to learn the solution outside of the boundary training data. After training we use later time steps from FD simulation to test the generalisation ability of the network at these later times.

For the last case study, our network is extended by concatenating the source location, $s$, to its inputs (shown in Figure~\ref{fig:model}), and it is trained to simulate the wavefield over many different source locations. In order to sufficiently constrain the network solution in this case we run FD simulations with many different initial source locations in the boundary training set, and randomly sample over the source location when computing the physics loss.

To train all networks we modify the loss function given by Equation~\ref{eq:rassi_loss} to
\begin{multline}
\label{eq:loss}
\mathcal{L}={1 \over N_{u}} \sum^{N_{u}}_{i=1}\|u_{\mathrm{FD}}(t_{i},x_{i},s_{i})-\Lambda(t_{i},x_{i},s_{i})\|^{2}+ \\
{k \over N_{\Lambda}} \sum^{N_{\Lambda}}_{j=1} \|\mathcal{N}[\Lambda(t_{j},x_{j},s_{j};\theta);v(x_{j})]\|~,
\end{multline}
where $v(x)$ is the fixed velocity model and $k \in \mathbbm{R}^{+}$ is a hyperparameter added to allow more control over the balance between the two terms. In testing we find that a L1 norm on the physics term shows better convergence than an L2 norm for this task (discussed further in Section~\ref{sec:marmousi}). For simplicity we only use starting time steps after the source term becomes negligible in the boundary data, and use a constant density model, which allows the differential operator $\mathcal{N} = \nabla^{2} - (1 / v^{2} ) {\partial_{tt} }$ to be used in the physics loss in Equation~\ref{eq:loss}. We write our own code to analytically compute the second order derivatives of the network in order to compute the physics loss, although auto-differentiation could also be used.

Curriculum learning is used to help the network converge to a solution. We start by training the network to reconstruct the boundary data only by using the boundary loss, and then \say{switch on} the physics loss halfway through training. This allows the network to learn the initial wavefield before learning to solve the wave equation through time. Furthermore a linearly growing time horizon is used when feeding points to the physics loss, analogous to the way a numerical solver iteratively solves the wave equation through time. Both schemes are intended to allow  the network to learn incrementally rather than all at once and in testing they are found to improve convergence (see Appendix~\ref{appendix:a}).

The network is trained using the Adam stochastic gradient descent algorithm \citep{Kingma2014}. For each update step a random set of discretised points are sampled from the initial wavefield to compute the boundary loss and a random set of continuous points over the full input space up to the current horizon time are sampled to compute the physics loss. A batch size of 500 samples in both sets is used for each update step with a learning rate of 1x$10^{-5}$.

\subsection{Boundary data generation}

We solve the wave equation using three different velocity models to test our approach, shown in Figure~\ref{fig:velocity}. The PINN is retrained in each setting and the models are chosen in increasing complexity to test the limits of the approach. They consist of a homogeneous velocity model with a fixed value of $2500~\mathrm{ms}^{-1}$, a horizontally layered velocity model and a region of the Earth-realistic Marmousi P-wave velocity model \citep{Martin2006}.  2D Gaussian smoothing with a standard deviation of 2 grid points is applied to the each model before it is used to alleviate discontinuity issues (discussed in more detail in Section~\ref{sec:results} and Appendix~\ref{appendix:a}).

We use the SEISMIC\_CPML FD modelling code \citep{Komatitsch2007} to generate the initial wavefield training data for each case. All simulations use a $300\times300$ grid with a spacing of 5~m in each direction (i.e. $X_\mathrm{max}=1500$~m in both directions), a $0.002$~s time interval, a constant density model of $2200~\mathrm{kgm}^{-2}$ and a 20~Hz Ricker source time function. For the first two velocity models a single simulation is used. The point source is placed in the centre of the velocity model and the first 10 time steps (i.e. $T_{1}=0.02$~s or 0.14 source cycles) after the source term becomes negligible are taken as training data. For the Marmousi velocity model 100 simulations are used each with a random source location drawn from a 2D Gaussian distribution located towards the bottom of the model (shown in Figure~\ref{fig:velocity}), and 20 time steps (i.e. $T_{1}=0.04$~s or 0.28 source cycles) are taken. For all cases the points used for computing the physics loss are sampled up to a maximum time of $T_\mathrm{max}=0.4$~s.

\section{\label{sec:results}Results}

\begin{figure*}[t]
\begin{center}
\includegraphics[width=16cm]{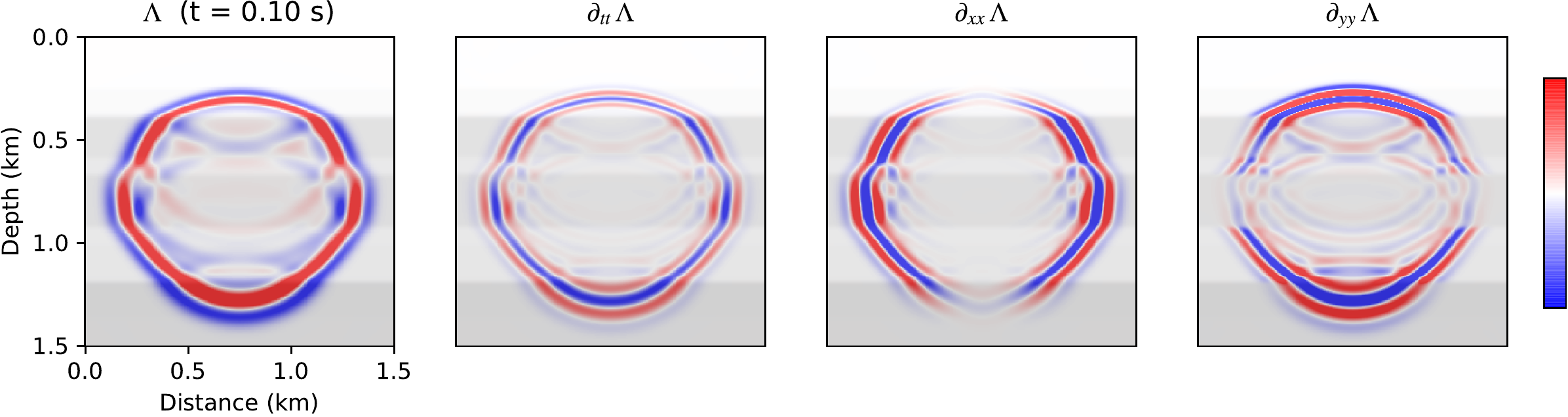}
\caption[]{Second order derivatives of the PINN trained using the layered velocity model. The derivatives are shown at a single snapshot in time and the plots are overlain on the velocity model. From left to right: network wavefield prediction, its second order time, second order $x$ and second order $y$ derivatives. Sharp changes occur in the second order $y$ derivative at velocity interfaces. Arbitrary scale used for colour bar.}
\label{fig:gradients}
\end{center}
\end{figure*}

\begin{figure*}[t]
\begin{center}
\includegraphics[width=14cm]{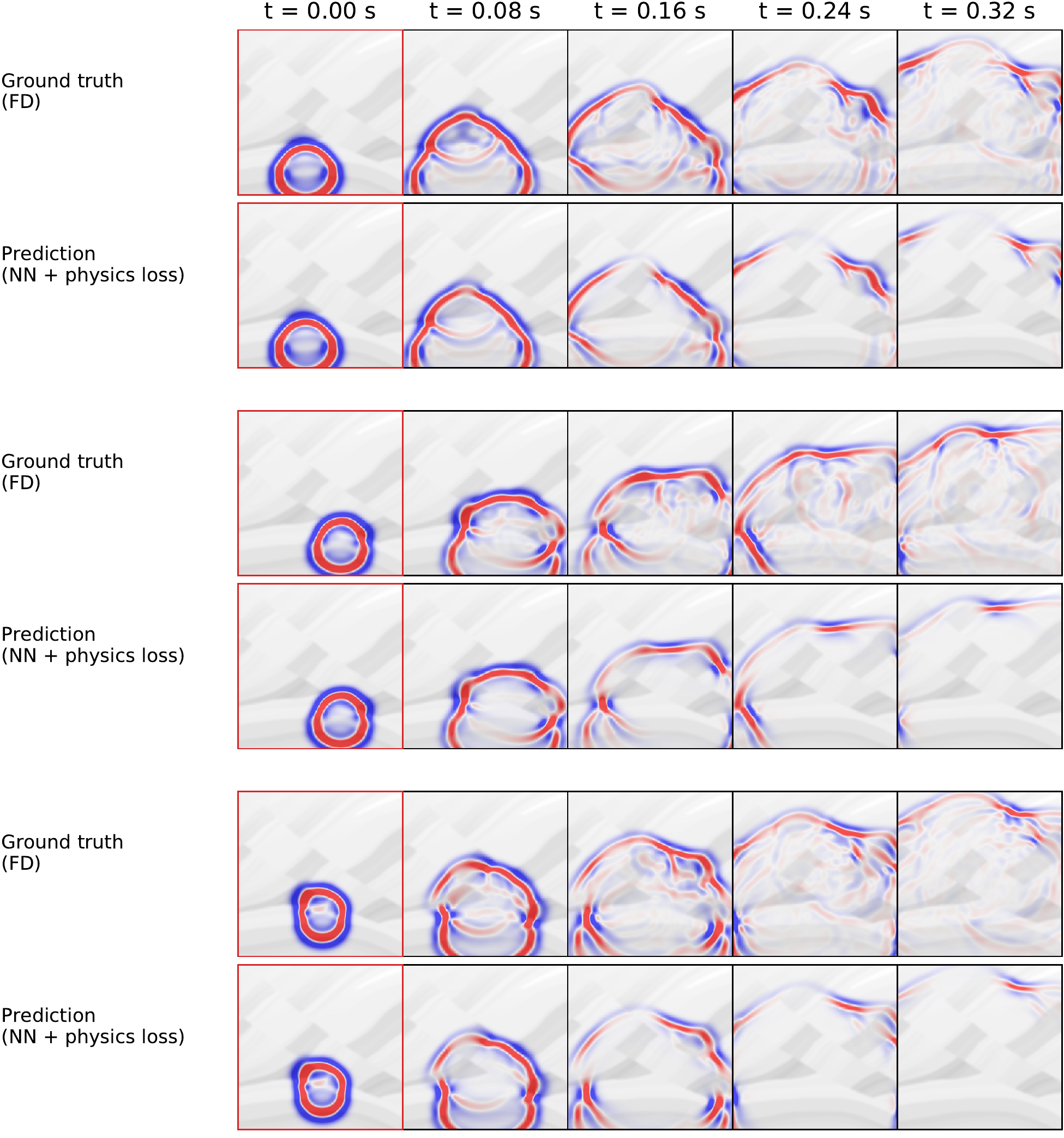}
\caption[]{Comparison of the PINN wavefield prediction to ground truth FD simulation, using the Marmousi velocity model. Each pair of rows shows the PINN prediction and FD simulation for a randomly selected source location in the test set. The plots are overlain on the velocity model, shown in grey. For this case the boundary training data only covers the time range 0.00-0.04~s and the PINN is able to generalise to much later times (beyond 0.32~s, 8$\times$ the range of the boundary data).}
\label{fig:marmousi}
\end{center}
\end{figure*}

\subsection{\label{sec:constant}Homogeneous velocity model, single source}

Our first case study tests the ability of the PINN in the simplest possible case: a homogeneous velocity model. In this media the initial wavefield propagates outwards with a fixed velocity and its amplitude reduces due to spherical divergence. We train the network and find that both the boundary loss and physics loss converge, and predictions of the wavefield after training are shown in Figure~\ref{fig:constant}. The network agrees with the FD solution very accurately, capturing both its kinematics and its amplitude attenuation. We also train the same network with only the boundary loss, and show its wavefield predictions in Figure~\ref{fig:constant}. In contrast to this network, the physics-informed network is able to accurately predict the wavefield at times much later than the initial wavefield, showing much better generalisation capability outside of the boundary training data. Despite its poor accuracy, the uniformed network is still able to capture the notion of the wavefront propagating outwards outside of the training data. This case study confirms the validity of the approach and gives us confidence to move onto the more complex case studies below.

\subsection{\label{sec:layer}Layered velocity model, single source}

Our second case study uses the horizontally layered velocity model, which induces much more complex wavefield dynamics; reflected and transmitted waves form at each velocity interface and the wavefield is compressed and expanded when travelling through different regions of the velocity model. The boundary and physics loss converge and the predictions of the PINN after training are shown in Figure~\ref{fig:layer}. The network is able to accurately capture the full range of dynamics: it is able to simulate transmitted and reflected waves which are not present in the initial wavefield, in contrast to the same network just trained using the boundary loss. This suggests that it is able to capture the specific physics phenomena occurring at the interfaces of the media. It is also able to accurately capture the compression, expansion and spherical spreading attenuation of the wavefield.

For this case we plot the boundary and physics loss during training in Figure~\ref{fig:loss}. The boundary loss slightly increases when the physics loss is added and then remains stable. The physics loss is relatively constant throughout training, even as later times are added in the expanding time horizon, suggesting on average it is kept at the same low level as the initial wavefield. However, looking in more detail at Figure~\ref{fig:layer} we observe that the loss is higher along the interfaces in the velocity model. This is perhaps due to the discontinuities in the velocity model which translate into discontinuities in the second order gradients of the wavefield along the $y$-direction, which the network may struggle to represent. We plot the second order gradients of the network in Figure~\ref{fig:gradients} and find that the network has learnt sharp, although not fully discontinuous, contrasts at the interfaces.

\subsection{\label{sec:marmousi}Marmousi model, multiple sources}

Our final case study uses the Marmousi velocity model. This is a complex, Earth-realistic model which includes faults, velocity inversions and a large range of dips (angles of interfaces). For this case study we further increase the complexity of the problem by conditioning the network on the source location, using the source locations shown in Figure~\ref{fig:velocity} to train the network. To test the network we generate a separate test set of 20 input source locations which are not used during training. We find that the boundary and physics loss converge, albeit at a slower rate than for the homogeneous and layered velocity models, and the predictions of the PINN for three of the source locations in this test set are compared to FD simulation in Figure~\ref{fig:marmousi}.

In this case the network is able to accurately simulate the kinematics of the initial wavefront throughout the media to times much later than the initial wavefield. The network is also able to generalise across the source position, accurately modelling the initial wavefield and subsequent dynamics without needing to be retrained. However, the network struggles to learn the secondary reflected waves of the initial wavefront as it propagates through the model. The reason for this is unclear, and one explanation could be that because of their small relative amplitudes these reflections do not make a large enough contribution to the loss function in Equation~\ref{eq:loss} to be modelled accurately. In testing we found that an L1 norm on the physics loss improved the accuracy of the network compared to an L2 loss, perhaps by increasing the contribution of lower wavefield amplitudes, and it is possible that further amplitude balancing is required. Another explanation is that there are also inaccuracies in the FD simulation. We discuss this issue in more detail in Section~\ref{sec:discussion}.

For this case study the PINN takes approximately 1 day to train using a single Nvidia Titan V GPU, although we make little effort to optimise its training time. The majority of this time is spent computing the second order derivatives of the network, which are approximately $10\times$ as slow to calculate as the forward pass of the network. After training, the network is very efficient when computing the wavefield for arbitrary source positions at arbitrary points in time and space (of the order of 0.001~s), whilst 2D FD modelling of the full media is much slower for each source simulation (of the order of 1~s, or 1000$\times$ slower).

\section{\label{sec:discussion}Discussion}

We have shown that physics-informed neural networks are able to accurately solve the wave equation in increasingly complex media. They are able to capture a large range of physics, including the transmission and reflection of waves at interfaces, wavefield compression and expansion through different velocity regions and spherical spreading attenuation. Impressively they are able to accurately model the wavefield dynamics at much later times than the boundary data used to train them, and are therefore a promising approach for addressing the generality issues in existing deep learning approaches.

We extended the original approach proposed by \cite{Raissi2019} by conditioning our PINN on source location, which is practically desirable as it allows multiple simulations to be carried out without needing to retrain the network. A potential application of this work could be for seismic hazards assessment, where the quantification of hazards in a given media across different source locations is desired \citep{Reiter1990, Jena2020}. Our method could provide a much more efficient approach than using multiple FD simulations for this task, both because the network does not need to be retrained and because the network can query the wavefield at specific points in time and space (rather than needing to solve for the entire wavefield). Our network could also be used for source inversion, as its gradients with respect to its inputs are readily accessible, and it is fully flexible to where the observed wavefield measurements are located. Another benefit of this approach is that the physics loss can be used to quantify the accuracy of the network's solution, which could be useful to understand its reliability.

A challenge identified by this work is the modelling of waves at interfaces, where they introduce discrete changes in the second order derivatives of the wavefield. We note that modelling these discontinuities is also a challenge for numerical methods, where oversampling is typically required to ensure the solution remains accurate \citep{Igel2017}. The performance of our network using the layered velocity model degraded at these interfaces and we somewhat avoided the problem by smoothing the velocity model before use, although errors were still observed along the interfaces. We show the network prediction when the layered velocity model without smoothing was used in Figure~\ref{fig:other_tests}. The accuracy of this solution was significantly worse than in Figure~\ref{fig:layer}. Another related issue was that the Marmousi test struggled to model reflected waves with low relative amplitudes. Future work could investigate these issues, for example by forcing the network to be discontinuous (e.g. by using separate networks for each velocity section), by changing the activation function \citep{Sitzmann2020}, by using the specific physics constraints (such as wavefield continuity \citep{Aki1980}) that arise at interfaces in the loss function, or by using a more intelligent sampling scheme, such as adaptive co-location \citep{Anitescu2019}, instead of random sampling, to sample interfaces more often in the loss function.

Another extension of this work is to condition the network on further initial conditions, such as the velocity model, as well as the source location. This would allow it to become even more general and potentially be useful for a wider range of tasks. Our last case study showed that our network has a high representational capacity and is able to model the initial wavefield and subsequent dynamics for many different sources, and it may therefore be able to generalise further. One approach may be to use an \say{encoder} network to reduce the velocity model into a set of salient features before feeding it to the fully connected network to help the network converge.

Finally, our approach is readily extendable to 3D and more complex wave equations. 3D extension is trivial in the network design, as the dimensionality of the input tensor just needs to be increased by 1. Furthermore it may not be too practically challenging to generate 3D training data, as our approach only requires the first few time steps of FD simulation to train the network. One speed up of our approach could be to simulate multiple sources in each FD simulation and separate their wavefields before using them as training data, to reduce the number of FD simulations required. For more complex wave equations, such as the elastic wave equation, the underlying physics loss in Equation~\ref{eq:loss} just needs to be modified. The performance of our PINN on the wave equation suggests that PINNs may perform well for other hyperbolic systems too. Another exciting possibility is to investigate how efficiently the network solves high-frequency simulations, which take considerably more computation using numerical methods. Our future work will investigate these directions.

\section{\label{sec:conclusion}Conclusion}

We have shown that physics-informed deep neural networks are able to solve the wave equation and generalise outside of their training set by adding physics constraints directly to their loss function. This offers a powerful new approach for simulating the wavefield which is efficient once trained and does not rely on discretisation. We further condition our network on source location, allowing it to predict the wavefield across many different source locations without needing to be retrained. Future work includes understanding the practical applications of this approach and extending it to solve more complex, 3D wave equations.

\begin{acknowledgements}

The authors would like to thank the Computational Infrastructure for Geodynamics (\href{https://www.geodynamics.org/}{www.geodynamics.org}) for releasing the open-source SEISMIC\_CPML FD modelling libraries.

\end{acknowledgements}

\bibliographystyle{plainnat}
\bibliography{main}

\appendix

\setcounter{table}{0}
\renewcommand{\thetable}{A\arabic{table}}

\setcounter{figure}{0}
\renewcommand{\thefigure}{A\arabic{figure}}

\section{\textbf{Supplementary figures}} \label{appendix:a}

\begin{figure}[!h]
\begin{center}
\includegraphics[width=5cm]{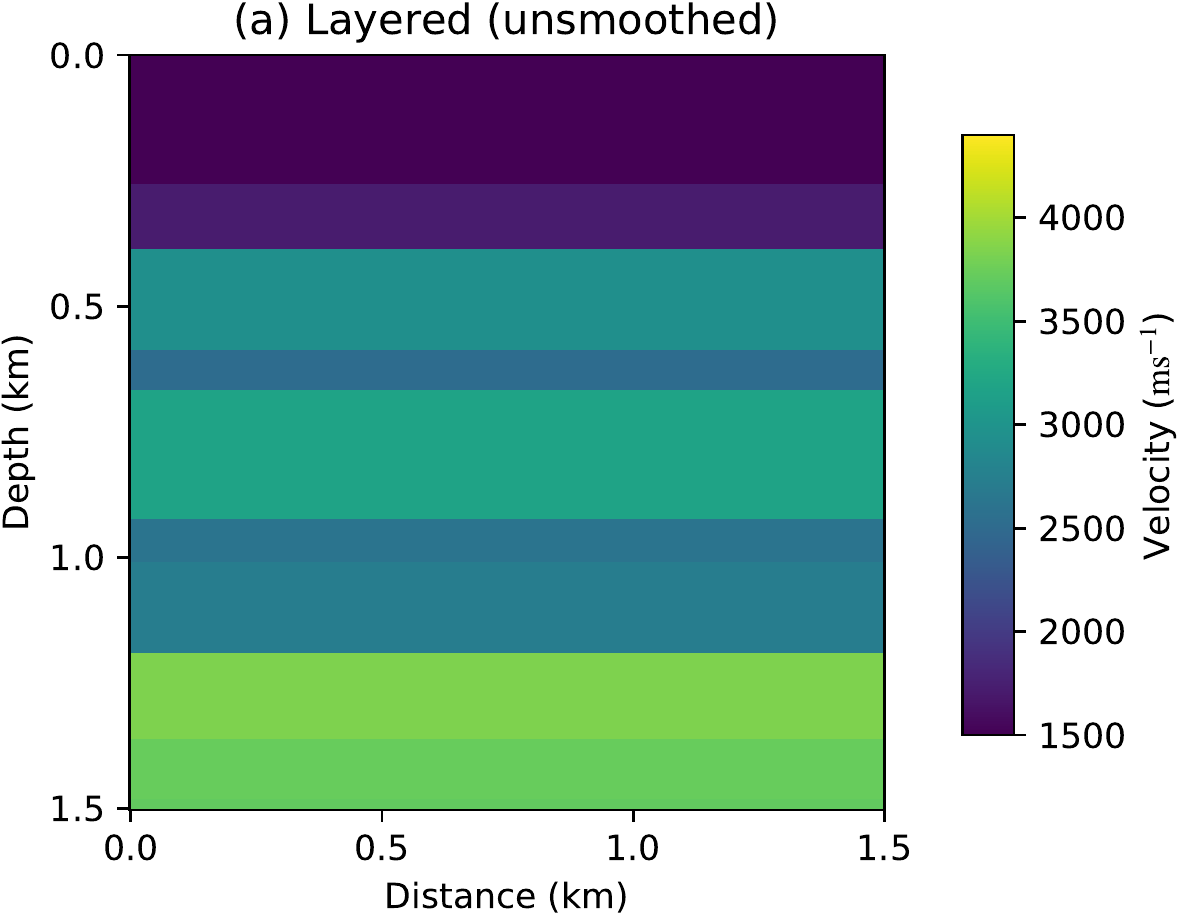}
\caption[]{Layered velocity model without smoothing, used in Figure~\ref{fig:other_tests}.}
\label{fig:velocity_unsmoothed}
\end{center}
\end{figure}

\begin{figure*}[t]
\begin{center}
\includegraphics[width=14cm]{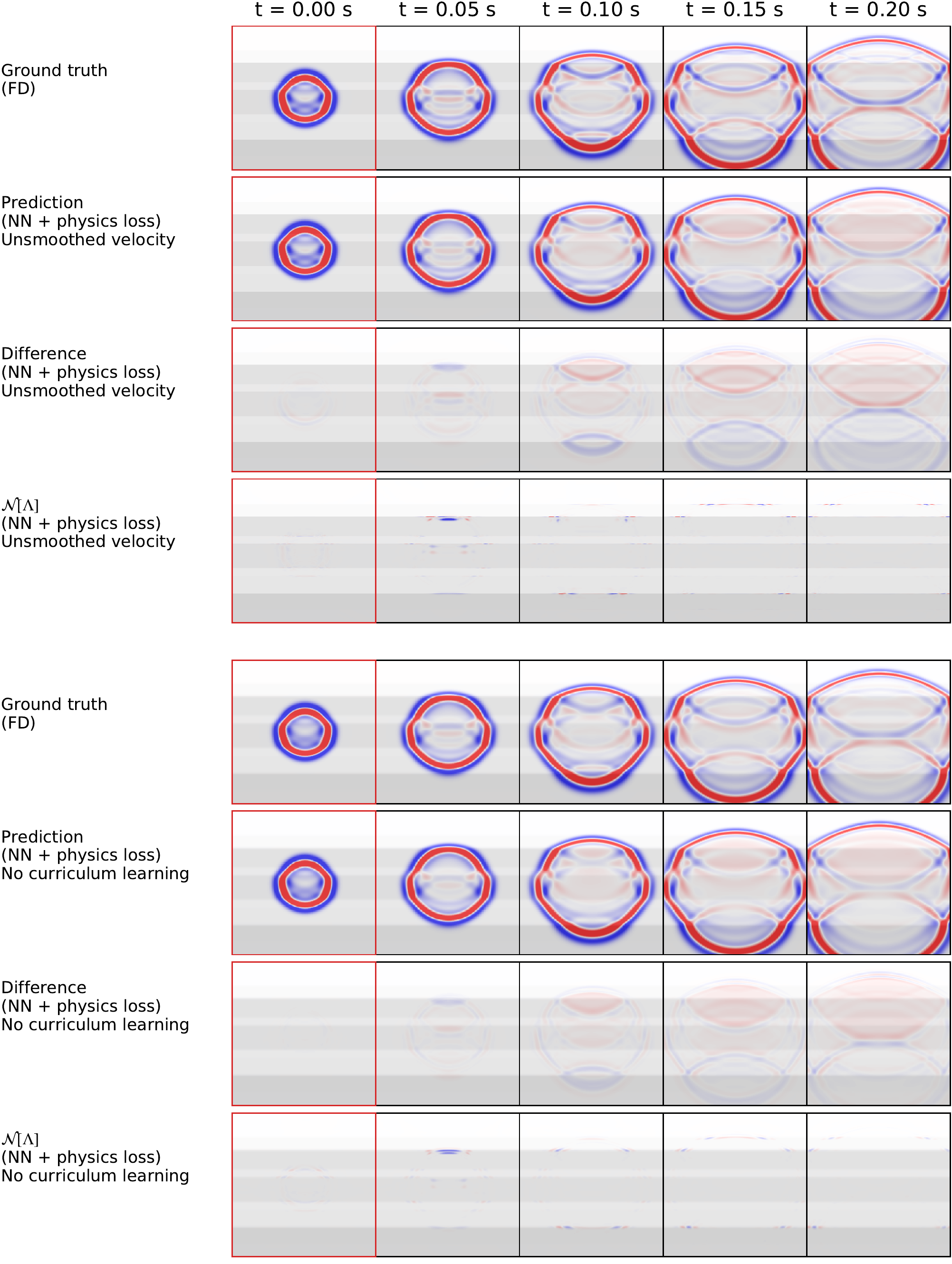}
\caption[]{Comparison of the PINN wavefield prediction to ground truth FD simulation for two additional PINN training setups. In the first test the layered velocity model is used without smoothing (\say{Unsmooothed velocity}), shown in Figure~\ref{fig:velocity_unsmoothed}. In the second test curriculum learning is not used; that is, the physics loss is used from the start of training and an expanding time horizon is not used (i.e. $t_{j}$ values for computing the physics loss in Equation~\ref{eq:loss} are sampled from their full input range $[0,T_{\mathrm{max}}$] from the start of training). Top four rows show the results from the first test and bottom four rows show the results from the second test. For each test, top two rows show the FD and PINN wavefield solutions through time. Third row shows the difference of the network prediction to FD simulation. Fourth row shows the value of the differential operator in the physics loss in Equation~\ref{eq:loss} (right hand term), which is close to zero for an accurate solution to the wave equation. Plots bordered in black indicate wavefield times which are outside of the boundary training data. Plots are overlain on the velocity model, shown in grey. Larger errors can be seen in both cases compared to the PINN in  Figure~\ref{fig:layer}.}
\label{fig:other_tests}
\end{center}
\end{figure*}

\end{document}